\documentclass[aps,twocolumn,10pt,nofootinbib]{revtex4-1}
\usepackage{graphicx,amssymb,amsmath}

\newcommand{\ket}[1]{\left|#1\right>}

\newcommand{\braket}[1]{\left<#1\right>}
\newcommand{\para}[1]{\left(#1\right)}

\begin{document}
\title{Phase diagram of the Hubbard model on honeycomb lattice}
\author{Abolhassan Vaezi and Xiao-Gang Wen}
\affiliation{Department of Physics, Massachusetts Institute of Technology,
Cambridge, MA 02139, USA}

\date{\today}

\begin{abstract} In this paper we generalized the slave-particle technique to
study the phase diagram of the Hubbard model on honeycomb lattice which may
contain charge fluctuations.  For large $U$, we have antiferromagnetic order
phase.  As we decrease  $U$ below $U_{c2}\simeq 3t$, the system undergoes a first
order phase transition into a gapped spin rotation invariant phase.  Under a
semiclassical approximation of the slave-particle approach, we find that such
phase breaks the translation symmetry, the parity and the lattice  rotation symmetry.
However, beyond the semiclassical approximation, a $Z_2$ spin liquid that does
not break any lattice symmetry is also possible.
\end{abstract}
\maketitle

{\em Introduction.---}
Hubbard
model\cite{Hubbard_1963,Hirsch_1985a} is believed to
describe the physics of many strongly correlated systems
e.g., Mott insulator\cite{Mott_1949,Imada_et_al_1998} and
high temperature
superconductors\cite{Bednorz_Mueller_1986,Anderson_1987Sci,Lee_Nagaosa_Wen_2006a}.
It is the simplest model one can write capturing the strong
correlation physics.
So far
many theoretical\cite{Hirsch_1983,Lieb_1989} and numerical
techniques\cite{Meng_2010a,Furukawa_1992,Lilly_1990} have
been developed to study this model. Among them is the slave
particle\cite{Anderson_Zou_a,Florens_Georges_2004a,PA_Lee_SS_Lee_2004a,Senthil_2008_a}
theory which was motivated by the RVB state first introduced by P.W. Anderson\cite{Anderson_1973}. One of the
interesting phases that have been studied and is strongly
supported by the slave particle approach is the $Z_2$ spin liquid
phase\cite{RS9173,W9164,Wen_2002c} which does not show any long range
order down to zero temperature. Unfortunately this phase has
not been experimentally verified but recently Meng {\em et
al}\cite{Meng_2010a} have studied the Hubbard model on
honeycomb lattice at half filling using the quantum Monte
Carlo (QMC) method and have reported the existence of a spin liquid
phase for a range of $U/t$. Fortunately QMC does not have
sign problem on bipartite lattices at half filling so we can
trust its results. For small U-limit they have reported the
semi-metallic phase. At $U_{c1}\sim 3.5t$ they have seen a
phase transition to the spin liquid with nonzero spin
excitation gap (gapped spin liquid). At $U_{c1}$ the charge
gap opens up and therefore this transition point is associated
with the Mott metal-insulator transition. For a larger value of $U_{c2} \sim
4.3 t$ they have obtained the anti-Ferromagnetic (AF) order in which the charge
gap is still nonzero but the spin excitation is the gapless Goldstone mode.

\begin{figure} \includegraphics[width=220pt]{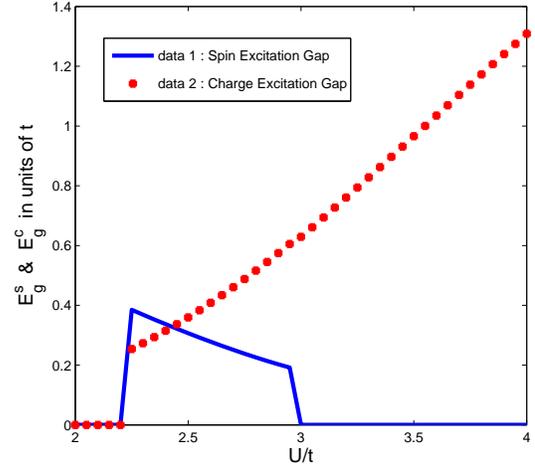}\\
\caption{ Spin excitation gap (blue line) and charge excitation gap (red dots)}\label{fig1} \end{figure}
\begin{figure} \includegraphics[width=220pt]{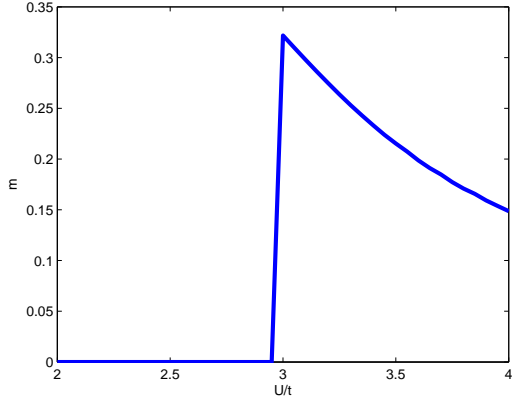}\\
\caption{Staggered magnetization, $m$
($(-)^{i}~S_z\para{i}$), as a function of $\frac{U}{t}$.
There is a phase transition to the antiferromagnetic order
at $U/t=3$. }\label{fig2} \end{figure}

In this paper, we generalized the slave-particle
method\cite{Anderson_Zou_a,Lee_Nagaosa_Wen_2006a} to capture charge
fluctuations\cite{FG,PA_Lee_SS_Lee_2004a} to study the Hubbard model on honeycomb lattice.  we
have obtained a similar phase diagram (see Fig.  \ref{fig1} and \ref{fig2}) as
in \cite{Meng_2010a} but with different numerical values for $U_{c1}$ and
$U_{c2}$.  We obtained a superconducting phase (instead of the semi-metal
phase) for small $U/t$ and a AF phase for large $U/t$.  At the meanfield level,
our phase between $U_{c1}$ and $U_{c2}$ is a spin liquid with finite
charge/spin gap that do not break any symmetry.  However, the meanfield state
is unstable.  Under a semiclassical approximation, we show that phase between
$U_{c1}$ and $U_{c2}$ is a charge/spin gapped state that breaks translation and
lattice rotation symmetry but not spin rotation symmetry.  On the other hand,
in the presence of the second neighbor hopping, the meanfield state may become
a $Z_2$ spin liquid state\cite{RS9173,W9164,Wen_2002c} that does not break any
symmetry and has finite charge/spin gaps.  All phase transitions are first
order which agrees with experiments\cite{Imada_et_al_1998}.

We would like to point out that the slave-rotor method is the other method to
include charge fluctuations,\cite{FG} which give rise to a nodal spin liquid
between $1.68 t< U <1.74 t$. The slave-rotor method is more reliable for small
$U/t$ and gives rise to the correct semi-metal phase.  Our method is quite
unreliable at small $U/t$ and gives rise to a (wrong) superconducting state.

In the large $U/t$ limit, the Hubbard model can be
approximated by the Heisenberg model and we expect strong AF
order in it. This model has been extensively studied by
different methods \cite{Jafari,Wang_Fa_2010a,Ying_2010_1,Sachdev_1989_1,Sachdev_1990_1,Fouet_1,Cenke_1}. Here we use a different approach to study the antiferromagnetic phase. It is shown that the spin/chrage gapped phase has an instability towards antiferromagnetism.


The Hubbard model is defined as the following:
\begin{eqnarray}
H=U\sum_in_{i,\uparrow}n_{i,\downarrow}-t\sum_{\left<i,j\right>,\sigma}C_{j\sigma}^\dag
C_{i\sigma}+h.c.  \end{eqnarray}

Here $\braket{i,j}$ means site $j$ is one of the nearest
neighbors of site $i$. We know that Hilbert space of Hubbard
Hamiltonian has four sates per site.
$\ket{0_f}_i,\ket{\uparrow}_i,\ket{\downarrow}_i,
\ket{\uparrow \downarrow}_i$. Let's name each state as
follows: $\ket{\mbox{holon}}_i=h_i^\dag \ket{\mbox{vac}}_{i}
= \left|0\right>_i,$ $\left|\mbox{spin up
spinon}\right>_i=f_{i,\uparrow}^\dag
\ket{\mbox{vac}}_{i}=\left|\uparrow\right>_i,$
$\ket{\mbox{spin down spinon}}_i=f_{i,\downarrow}^\dag
\ket{\mbox{vac}}_{i}=\ket{\downarrow}_i,$ $
\ket{\mbox{doublon}}_i=d_{i}^\dag\ket{\mbox{vac}}_{i}=\ket{\uparrow
\downarrow}_i$ in which $\ket{\mbox{vac}}$ is the vacuum, an
unphysical state which contains no slave particles even
holons. Using this picture we can rewrite the electron
creation operator as: $C_{i,\sigma}^\dag= f_{i,\sigma}^\dag
h_{i}+\sigma ~ d_{i}^\dag f_{i,-\sigma}=[\begin{array}{cc}
h_i & d_i^\dag \end{array} ] \left[ \begin{array}{c}
f_{i,\sigma}^\dag \\ \sigma f_{i,-\sigma} \end{array}\right]
$.

It should be mentioned that the physical Hilbert space
contains only four states: empty state (holon), one electron
(spinon) and two electrons (doublon) on each site. So we
always have one and only one slave particle on each site. So
we conclude that we should put the local constraint
$n_{i}^{h}+n_{i,\uparrow}^{f}+n_{i,\downarrow}^{f}+n_{i}^{d}=1$,
to get rid of redundant states. This is the physical
constraint which should be satisfied on every site. We could
also obtain this result by noting that the electron
operators are fermion and should satisfy the anticommutation
relations. From the definition of $ C_{i,\sigma}^\dag$ it is
obvious that it is invariant under U(1) gauge transformation
(We require $h_{i}$ and $d_{i}$  to remain bosonic operators
{\em i.e.}, preserve their statistics after transformation,
otherwise we would have SU(2) gauge invariance. However at
$U=\infty$ we have only fermions and only in that case we
have SU(2) gauge symmetry). It is worth noting that all
the slave particles carry the same charge under the internal
U(1) gauge. Since the constraint as well as the
Hubbard Hamiltonian are gauge invariant, so is the
action of the Hubbard model.

In terms of new slave particles, the Hubbard Hamiltonian can
be written as:
\begin{eqnarray}
&&H=\sum U d_{i}^\dag
d_{i}-t\sum_{\braket{i,j}}\para{\chi_{i,j}^{f}\chi_{j,i}^{b}+\Delta_{i,j}^{f
\dag}\Delta_{i,j}^{b }+h.c.}~~~~
\end{eqnarray}
In which we
have used these notations
$\chi_{i,j}^{f}=\sum_{\sigma}f_{i,\sigma}^\dag
f_{j,\sigma}~,~\chi_{i,j}^{b}=h_{i}^\dag h_{j}-d_{i}^\dag
d_{j}~,~\Delta_{i,j}^{f}=\sum_{\sigma}\sigma f_{-\sigma,i}
f_{j,\sigma}~,~\Delta_{i,j}^{b}= d_{i}h_{j}+h_{i}d_{j}$ . To
implement the constraint we appeal to the path integral and
the Lagrange multiplier methods.
\begin{eqnarray}
&&S=\int
D f^\dag Df D h^\dag Dh D d^\dag Dd ~e^{-\int d\tau L }\cr
&& L=f_{i,\sigma}^\dag \frac{\partial }{\partial
\tau}f_{i,\sigma}+ d_{i}^\dag \frac{\partial }{\partial
\tau}d_{i}+h_{i}^\dag \frac{\partial }{\partial
\tau}h_{i}+i\lambda_{i}g_{i} +H \cr
&&g_{i}=f_{i,\uparrow}^\dag f_{i,\uparrow}
+f_{i,\downarrow}^\dag f_{i,\downarrow}+h_i^\dag
h_i+d_i^\dag d_i -1
\end{eqnarray}
The above motivates us to
define the effective Hamiltonian as
$H_{eff}=H+i\sum_{i}\lambda_{i} g_{i}$. Now by using the
Hubbard-Stratonovic transformation we can decouple spinons
from [hard-core] bosons at the mean field level. To do so we
just replace $\chi_{i,j}$ and other operators with their
average. For translational invariant systems we can assume:
$\braket{\chi_{i,j}}=\braket{\chi_{i-j}}$ and so on. From
now on $\chi$ stands for the average of $\chi$ operators and
so on. Moreover $\braket{i\lambda_{i}}=\lambda_{0}$. By
these assumptions we can obtain unknown parameters in the
effective Hamiltonian from self-consistency equations. Now, let us focus on the effective Hamiltonian of bosons. As long as $\Delta_{f}$ is nonzero, the pairing between holons and doublons is nonzero, and they form bound state. Using the Bogoliubov transformation we can show that the ground-state wave-function of bosons is a paired state which is
completely symmetric between holons and doublons. Therefore, as
long as this state represents the ground, we have $\braket{h_{k,A}^\dag h_{k,B}}=\braket{d_{k,A}^\dag d_{k,B}}$, and as a result:
$\chi_{b}=\braket{h_{i,A}^\dag h_{j,B}-d_{i,A}^\dag d_{j,B}}=0$.
Spinons cannot hop in this case and the system is insulator. Self-consistent equations show that $\chi_{f}=0$ as well and therefore the following Hamiltonians describe the low energy theory of this phase:




\begin{widetext}
\begin{eqnarray} &&H_{f}^{A,B}=\sum_{k} \left[
\begin{array}{cc} f_{k,A,\uparrow}^\dag & f_{-k,B,\downarrow} \\
\end{array} \right]\left[ \begin{array}{cc}
-\lambda_{0}~~~&~~~ -t\Delta_{k}^{b} \\
-t\Delta_{k}^{b} ~~~&~~~ +\lambda_{0} \\
\end{array}\right] \left[ \begin{array}{c} f_{k,A,\uparrow} \\
f_{-k,B,\downarrow}^\dag \\ \end{array} \right]\\
&&H_{b}^{A,B}=\sum_{k} \left[ \begin{array}{cc} ~d_{k,A}^\dag~~ &
h_{-k,B}~ \\ \end{array} \right]\left[ \begin{array}{cc}
U-\lambda_{0}~&~ -t\Delta_{k}^{f} \\
-t\Delta_{k}^{f} ~&~ -\lambda_{0} \\
\end{array}\right] \left[ \begin{array}{c} ~d_{k,A}~ \\
~h_{-k,B}^\dag~ \\ \end{array} \right]
\end{eqnarray}
\end{widetext}

where $\Delta^{f,b}\para{\vec{k}}=\sum_{\delta}\Delta_{\vec{\delta}}^{f,b}~e^{i\vec{k}.\vec{\delta}}$ and $\vec{\delta}$ connects two nearest neighbors. We have similar equations for $H_{f}^{B,A}$ and $H_{b}^{B,A}$. Using the Bogoliubov transformation we can diagonalize the
above Hamiltonians. The energy eigenvalues for spinons are
$E_{k}^{f}=\sqrt{\lambda_{0}^2+\left(t\Delta_{k}^{b}\right)^2}$.
For bosonic part we obtain: $E_{b}^{\pm,k}=
+\pm \frac{U}{2}+
\sqrt{\left(\frac{U-2\lambda_{0}}{2}\right)^2-\left(t\Delta_{k}^{f}\right)^2}$. At half filling, in order to excite a charge we need to annihilate
two spinons and create a pair of holon-doublon. So we can
define the charge excitation gap as the sum of the
excitation energy of a holon and a doublon. When the charge
gap is nonzero then the paired holon-doublon state is
stable because exciting quasi-particles on top of this state costs energy. For this state, the charge gap is $E_{c}^{g}=$ min$E_{b}^{+,k}$+min$E_{b}^{-,k}$= $2\sqrt{\left(\frac{U-2\lambda_{0}}{2}\right)^2-\left(3t\Delta^{f}\right)^2}$. Therefore, as long as $U-2\lambda_{0} > 6t\Delta_{f}$, charge gap is finite and we are in the insulating phase.


On the other hand, when the charge gap closes, the
paired holon-doublon state becomes unstable and free holons
and doublons proliferate. In this state, doublons and holons condense independently (single boson condensation)
such that $\braket{d_{i,A}}=-\braket{d_{i,B}}$ and $\braket{h_{i,A}}=\braket{h_{i,B}}$, and
therefore $\chi_{b}=2\braket{h_{i,A}}^2=2n_{h}\neq 0$, therefore spinons can hop freely and the ground state is no longer an insulator. Since doublons condense at sublattice $A$ and $B$ with opposite signs, we show that $\Delta_{b}=0$ and as a result $\Delta_{f}=0$. we relate the onset
of single boson condensation, {\em i.e.} the critical point below which charge gap closes, to the Mott transition. It should be mentioned that $\chi_{f,b}$ as well as $\Delta_{b,f}$ jump at this point, so we obtain a first order phase
transition in this way, which is consistent with experiments. We like to point out that since $d^\dag h$ operator that carries $2e$ electric
charge, condenses in this state,
we indeed obtain a superconducting state instead
of a semi-metallic phase.




{\em Phase diagram.---} In the following sections we discuss
the three phases that we have obtained from the slave
particle
method.

{\em Superconducting phase.---} Now let us approach the Mott
transition point from below i.e. from superconducting side. In
this phase both $\Delta_f$ and $\Delta_b$ are zero and
therefore the charge excitation gap as well as the spin excitation gap vanishes. Gapless charge excitation implies : min$E_{h,k}^{b}$+min$E_{d,k}^{b}=U-2\lambda -6
\chi_{f}=0$.  This condition can be satisfied up to
$U_{c1}=2\lambda+6t\chi_{f}=2.2t$. At this point the Mott
transition happens. In terms of physical electrons, we obtain an s-wave superconducting state with gapless charge and spin excitations. The pairing order parameter changes sign under parity and 60 degrees rotation and transforms trivially under all other symmetry transformations. It should be mentioned that at small U limit, the Bose gas of holons and doublons becomes very dense and there is strong interaction between them. So the mean-field results are unreliable in this regime and the superconducting state is a fake result. However, our method captures two important right features of the system below the phase transition, because we obtain zero spin excitation energy as well as zero charge excitation energy.



{\em Charge/spin gapped phase.---} For $U>U_{c1}$ we have
$\chi_{b}=0$. So the quasi-particle weight of spinons are
zero and they cannot hop since for any $i$ and $j$ arbitrary
sites:  $\braket{f_{j,\sigma}^\dag f_{i,\sigma}}=0$.
Therefore this state is like a superconductor with infinite
carrier's mass $m \sim \frac{1}{t\chi_{b}}\rightarrow \infty
$. Now let us find $U_{c1}$. To do so we assume that:
$\Delta_{f,b}\para{\vec{\delta}}=\Delta_{f,b}$. So we have
$\Delta_{f,b}\para{\vec{k}}=\Delta_{f,b}\eta(\vec{k})$, where $\eta(\vec{k})=
e^{ik_y}+2e^{-i\frac{ky}{2}}cos{\frac{\sqrt{3}}{2}k_x}$ and
therefore the energy spectrum of spinons and bosons are
$\sqrt{\lambda^2+\left|t\Delta_{b}\para{k}\right|^2}$ and
$\pm
U/2+\sqrt{\para{U/2-\lambda}^2-\left|t\Delta_{f}\para{k}\right|^2}$
respectively. From the energy dispersion of bosons, one can
read that the charge gap closes when
$U_{c1}=2\lambda+6t\Delta_f$. Our numerical results show
that near the phase transition, $\Delta_{f}\simeq .5$ and
$\lambda \simeq -.4$ and the Mott transition occurs at
$U_{c1}/t=2.2$.  For large $U/t$ limit:
$\Delta_{f}\rightarrow .53$ , $\Delta_{b}\sim \frac{t}{U}$ ,
$\lambda \sim (\frac{t}{U})^3Ln\frac{t}{U}$  and  $n_b \sim
\Delta_{b}^2 \sim (\frac{t}{U})^2$. It is clear from the
energy spectrum of spinons that in the spin liquid phase,
there is a gap in their spectrum equal to:
$E_{g}^{f}=|\lambda|$. Note that in the spin-charge
separation picture, the physics of spin is determined by
that of spinons. Therefore the spin excitation gap is also
$E_{g}^{s}=|\lambda|$.

Now let us focus on the gauge theory of this phase. In this
phase the effective action of spinons is of the following
forms:

\begin{eqnarray} H_{s}=&&\lambda\sum_{i,\sigma,\tau}
f_{i,\tau,\sigma}^\dag f_{i,\tau,\sigma} \cr &&
-t\sum_{<i,j>,\sigma,\tau}\Delta_{b}\para{i,j}\sigma
f_{i,A,\sigma}^\dag f_{j,B,-\sigma}^\dag +h.c.
\end{eqnarray}

Now if we transform operators as: $f_{i,A,\sigma}\rightarrow
e^{i\alpha}f_{i,A,\sigma}$ and $f_{i,B,\sigma}\rightarrow
e^{-i\alpha}f_{i,B,\sigma}$ for any arbitrary phase
$\alpha$, {\em i.e.} assuming a staggered global gauge
transformation, then the effective Hamiltonian does not
change. Therefore the invariant gauge group ($IGG$) of the
Hamiltonian is the staggered $U(1)$. The reason is that
there is no hopping term due to the nonzero charge gap and
the gauge transformation of two neighboring sites
have opposite phases, the total phase change of the pairing
term becomes zero and therefore gauge fluctuations are
described by staggered compact $U(1)$ instead of compact
$U(1)$ gauge theory. This is equivalent to assuming
positive unit charge on sublattice A and negative unit
charge on sublattice B for slave particles under the
internal gauge transformation.

So, at mean field level, the charge/spin gapped phase has a
neutral spinless $U(1)$ gapless mode as its only low energy
excitations.  However, it is well known that $U(1)$ theory
in 2+1D is confined due to instanton effects.
So let us assume that the $U(1)$ fluctuations are weak
and use the semiclassical approach to study
the $U(1)$ confined phase where the $U(1)$ mode is gapped.
We find\cite{Vaezi_2010c} that these instanton operators,
$e^{i\theta}$ (in the dual XY model),
carry a non-trivial crystal momentum.  Also, under 60 degree
lattice rotation and parity, an instanton is changed to an
anti-instanton,
$e^{i\theta} \to e^{-i\theta}$.  The instantons carry trivial quantum
numbers for other symmetries.  However, a triple instanton operator $
\cos(3\theta)$
carries trivial quantum numbers for all symmetries.  This
allows us to conclude that the
neutral spinless $U(1)$ mode is described by
$L= \frac{1}{2g} (\partial \theta)^2 + K \cos(3\theta)$.
In the semiclassical limit (the small $g$ limit),
$\braket{ e^{i\theta} } \neq 0$ and we obtain a phase
that breaks the translation, the parity and the $60^\circ$ rotation symmetries,
but not spin rotation symmetry.

We like to point out that in the presence of second neighbor
hopping in the Hubbard model the charge/spin gapped phase
can be spin liquid that do not break translation, parity, 60 degree
lattice rotation, and  spin rotation symmetries.
It is because we can break the staggered compact $U(1)$ gauge symmetry down
to a $Z_2$ one by Anderson-Higgs mechanism. If we add second
neighbor hopping to the Hubbard model, within slave particle
approach, this term generates pairing terms of the form
$f_{i,\tau,\sigma}^\dag f_{j,\tau,-\sigma}^\dag$, {\em i.e.}
it induces the same sublattice pairing and the Hamiltonian
is no longer invariant under the staggered global $U(1)$
gauge transformation. In this case the staggered compact
$U(1)$ gauge symmetry is broken down to a $Z_2$ gauge
symmetry.  The $U(1)$ gauge fluctuations are gapped and thus
our mean filed state is stable and we can trust our
meanfield results. Therefore we obtain a spin liquid phase.

\noindent{\em Antiferromagnetic phase.---} In this part
we show that the charge/spin gapped phase is unstable towards
antiferromagnetic order above $U_{c2}=3t$. To obtain Neel
order in the t-J model we simply assume that
$\braket{\vec{S}_{z,A}}=-\braket{\vec{S}_{z,B}}=m$. But how
can one implement this idea in the Hubbard model within
slave particle approach? In the Neel order phase, translation
symmetry is broken and there is an asymmetric situation
between sublattice A and B. For example we can obtain a
antiferromagnetic phase by assuming
$\Delta_{1,f}=\braket{f_{j,B,\downarrow}f_{i,A,\uparrow}}\neq
\braket{f_{j,A,\downarrow}f_{i,B,\uparrow}}=\Delta_{2,f}$.
This assumption simply means that the chance of finding a
spin-up spinon on sublattice A and another spin-down spinon
on sublattice B is more than finding the opposite one, so
this method introduces staggered sublattice magnetization
and leads to the Neel order. If there is a Neel order in the
system then the chance of creating one holon-doublon pair
from annihilating a spin-up spinon on sublattice A and a
spin-down spinon on sublattice B is more than the other
process. Therefore the excitation energy of spinons for
up-spin on  A and down-spin on B is
$E_{f}^{1}(k)=\sqrt{\lambda^2+|t\Delta_{1,b}(k)|^2}$ while
for down-spin on A and up-spin on B is
$E_{f}^{2}(k)=\sqrt{\lambda^2+|t\Delta_{2,b}(k)|^2}$. On the
other hand, since we are not interested in CDW, we need a
symmetric situation between sublattices A and B for the
charge sector. So the energy excitation of bosons is
$E_{b}(k)=\sqrt{\para{U/2-\lambda}^2-|t\Delta_{f}(k)|^2}$.
Using these assumptions we lead to the following
self-consistency equations:

\begin{eqnarray} &&
\Delta_{1,f}=\frac{t}{N_s}\sum_{k}\frac{|\eta(k)|^2\Delta_{1,b}}{E_{1,f}(k)}\\
&&
\Delta_{2,f}=\frac{t}{N_s}\sum_{k}\frac{|\eta(k)|^2\Delta_{2,b}}{E_{2,f}(k)}\\
&& \Delta_{f}=\Delta_{1,f}+\Delta_{2,f}\\ &&
\frac{\Delta_{1,b}\Delta_{1,f}+\Delta_{2,b}\Delta_{2,f}}{\Delta_{1,f}+\Delta_{2,f}}=\frac{t}{N_s}\sum_{k}\frac{|\eta(k)|^2\Delta_{f}}{E_{b}(k)}
\end{eqnarray}

By solving the above equations we find that above $U_{c2}=3
t$, $m \neq 0$. So we conclude that for $U>U_{c2}$ we obtain
AF order. It is interesting that in this phase, the gap of
spinons is very small and negligible (for example at U=4, it
is $-2\times 10^{-7}$). So in this phase we can assume
that spinons are massless quasi-particles.

In conclusion, we have used a generalized slave particle method to derive the phase diagram of the Hubbard model at half filling on the honeycomb lattice. Within the mean field approximation we
can decouple fermions from bosons to achieve the effective
Hamiltonian that describes the low energy physics of the
system. The physics of the Mott transition
is discussed and it turns out to be a first order phase transition. It
is shown that the phase transition occurs when the
charge gap opens up. Above the critical point, within meanfield theory we obtain a spin liquid phase. But after including gauge
 fluctuations of the emergent spin liquid and investigating the instanton effect, we argue that this phase is unstable and we finally obtain a spin/charge gapped phase that breaks the translation symmetry. For large U limit, a new approach to study
antiferromagnetic phase within the slave particle picture has been developed.
It is shown that the gapped spin liquid phase has an
instability towards antiferromagnetism.

{\em Acknowledgement.---} We thank B. Swingle, T. Senthil, P.A. Lee, M.
Barkeshli and T. Grover for their useful comments and helpful discussions.
This research is partialy supported by  NSF Grant No.  DMR-1005541.


%

\end{document}